\documentclass[12pt]{article}
\usepackage[xdvi]{graphicx} 
\usepackage{epsfig}
\usepackage{amsfonts,amssymb,amsmath}

\newcommand{\al}{\ensuremath{\alpha}}
\newcommand{\ga}{\ensuremath{\gamma}}

\newcommand{\ka}{\ensuremath{\kappa}}
\newcommand{\la}{\ensuremath{\lambda}}
\newcommand{\La}{\ensuremath{\Lambda}}

\newcommand{\del}{\ensuremath{\partial}}
\newcommand{\Del}{\ensuremath{\nabla}}

\newcommand{\be}{\begin{equation}}
\newcommand{\ee}{\end{equation}}

\newcommand{\ba}{\begin{eqnarray}}
\newcommand{\ea}{\end{eqnarray}}
\newcommand{\sech}{\textrm{sech}}
\newcommand{\cosech}{\textrm{cosech}}
  
\begin{document}

\rightline{hep-th/0402079}
\rightline{OUTP-04/04P}
\vskip 1cm 

\begin{center}
{\Large \bf Ghost-free braneworld bigravity}
\end{center}
\vskip 0.5cm
 
\begin{center}
{\it Dedicated to the memory of Ian Kogan}
\end{center}
\vskip 1cm 
\renewcommand{\thefootnote}{\fnsymbol{footnote}}

\centerline{\bf Antonio Padilla\footnote{a.padilla1@physics.ox.ac.uk}}
\vskip .5cm

\centerline{\it Theoretical Physics, Department of Physics}
\centerline{\it University of Oxford, 1 Keble Road, Oxford, OX1 3NP,  UK}

\setcounter{footnote}{0} \renewcommand{\thefootnote}{\arabic{footnote}}
 

\begin{abstract}
We consider a generalisation of the DGP model, by adding a second
brane with localised curvature, and allowing for a bulk cosmological
constant and brane tensions. We study radion and
graviton fluctuations in detail, enabling us to check for ghosts and
tachyons. By tuning our parameters
accordingly,  we find bigravity models that are free from ghosts and
tachyons. These models will lead to large distance modifications of
gravity that could be observable in the near future.
\end{abstract}

\newpage
\section{Introduction}
Where does the force of gravity come from? Anyone unfortunate enough
to be near a strongly gravitating object, such as a black hole, would
say ``from the curvature of spacetime''. In a weaker gravitational
field, we measure fluctuations about some background spacetime~\cite{Fierz:pauli-fierz}. These
fluctuations correspond to spin-2 particles called
gravitons. Traditionally, we now say that ``gravity comes from the
exchange of {\it massless} gravitons''. We believe  this because it reproduces Newton's Law
of gravitation, and that is well tested experimentally. But how well
is Newton's Law really tested? The truth is that our experimental
knowledge of gravity only covers distances between $0.2$mm and
$10^{26}$cm. The lower bound is as a result of Cavendish experiments,
and the upper bound corresponds to 1\%  of the current Hubble
length. In units of $\bar h=c=1$, this is equivalent to an energy scale
\be \label{range}
10^{-31}~\textrm{eV} < p < 10^{-3} ~\textrm{eV}
\ee
Perhaps, therefore, we should alter our previous statement. Gravity at
the experimental scale (\ref{range}) could be due to the exchange of massless and/or massive particles,
so long as the masses are less than $10^{-31}~\textrm{eV}$.  Indeed,
massive gravitons appear automatically in some higher derivative
gravity theories~\cite{Hindawi:spin2}. The
simplest non-trivial scenario would contain a single massive
graviton~\cite{Fierz:pauli-fierz}. If we have a combination of gravitons of many
different masses,  we have {\it multigravity}~\cite{Kogan:2000,
 Kogan:branemulti, Kogan:modification, Kogan:6dmulti,Kogan:review,
Papazoglou:multithesis}. In this paper, we will encounter {\it
bigravity}~\cite{Kogan:adsbranes, Damour:univclass}. This is the
simplest example of multigravity,  in which gravity is mediated by
both a massless graviton and a single {\it ultralight} graviton.

Multigravity is clearly of
interest from a purely theoretical point of view. However, it is also
of interest to phenomenologists because it predicts new
gravitational  physics at very large distances. At distances beyond
the Compton wavelength of the ultralight mode, the ultralight mode is
turned off, and gravity is mediated by the massless mode alone. Large
distance modifications of gravity have been {\it in vogue} recently,
as they could offer an explanation to the current acceleration of the
universe~\cite{Perlmutter:acc, Riess:acc, Deffayet:DGPcosmo, Damour:nonlin,
Lue:dark,Lue:cosmic}.
 
Unfortunately, there are number of problems with many existing models
of modified gravity. One very serious problem is
the presence of ghosts (see, for example, ~\cite{Kogan:2000,
Gregory:GRS, Pilo:ghost, Dubovsky:ghost, Dubovsky:DGP}). Although the ``AdS
brane'' model in~\cite{Kogan:adsbranes} is ghost-free, modifications
of gravity are hidden behind the AdS horizon. The six dimensional
model of Kogan {\it et al}~\cite{Kogan:6dmulti} is also thought to be
ghost-free, although this has not been confirmed as the model
is too difficult to work with. The DGP
model~\cite{Dvali:DGPmodel}, meanwhile, is simple, ghost-free, and even predicts new infra-red physics, that could one day be
observable. However, it is {\it not} a bigravity model; gravity is due
to a resonance of continuum modes. 

Let us describe the DGP model in more detail. It is given by the
following action
\be \label{DGPaction}
S_\textrm{DGP}=M^3 \int_\textrm{bulk} \sqrt{g} R(g)+m_{pl}^2
\int_\textrm{brane} \sqrt{\gamma}R(\gamma)
\ee
In other words, we have a four-dimensional Minkowski brane embedded in
a five-dimensional Minkowski bulk. The key ingredient is the brane localised
curvature, $R(\ga)$. This could be generated by quantum
corrections, if matter were present on the brane. Localised curvature
can also appear in string theory~\cite{Corley:EH,Antoniadis:CY}. In
the DGP model, Newton's Law is
reproduced up to a distance $r=m_{pl}^2/2M^3$.  Beyond this scale the
behaviour of gravity is five-dimensional. 

In this paper, we will consider a generalisation of the DGP model. We
will add a second brane with localised curvature. We will also allow
for a bulk cosmological constant, and for the branes to have
tension. Since the extra dimension is finite, we will get
a discrete graviton mass spectrum. Our aim is to ask the following question: is it possible to
obtain an interesting modified theory of gravity {\it without}
introducing ghosts and tachyons? The answer will be ``yes''. For
certain parameter regions, we will discover bigravity models that are
tachyon and ghost-free, leading to potentially  observable new physics
in the infra-red. It is interesting to note that a single
DGP-like brane in a compact extra dimension also has a discrete mass
spectrum, but does {\it not} exhibit bigravity~\cite{Dvali:power}.

The rest of this paper will be organised as follows: in
section~\ref{sec:setup}, we will describe our set up in more detail,
giving the bulk and boundary equations of motion. In
section~\ref{sec:background}, we  will derive solutions for the
background spacetime. We will perturb about this background in
section~\ref{sec:pert}, arriving at the linearised equations
of motion. In section~\ref{sec:radion} we will focus on the {\it
radion} mode. This corresponds to fluctuations in the brane
separation. We will calculate its effective action to quadratic order, in
order to determine whether or not it is a ghost. For de Sitter branes
we will find that the radion is tachyonic. In
section~\ref{sec:spectrum}, we  will turn our attention to the
gravitons. We will derive the mass spectrum, and show that we can have
{\it observable} bigravity. We will also check for ghosts by
calculating the graviton effective action. In
section~\ref{sec:analysis}, we will play around with our parameters
until we find a bigravity model that is ghost-free, and might one day
lead to observable, new, infra-red
physics. Section~\ref{sec:conclusions} will contain some final
remarks. In particular, we will discuss some important issues, such as the famous van Dam-Veltman-Zakharov (vDVZ)
discontinuity~\cite{vanDam:VDVZ, Zakharov:VDVZ}, and
the recently discovered strong coupling
problems in massive gravity~\cite{Arkani-Hamed:massive,
Schwartz:decon}. Finally, the appendix contains a detailed calculation
of the mass spectrum for AdS branes.

\section{The Set Up} \label{sec:setup}

Consider  two $3$-branes embedded in a
five-dimensional bulk spacetime. For
simplicity we assume that there is $\mathbb{Z}_2$ symmetry across each
brane. This means that we can  focus on the manifold, $\mathcal{M}$,  which has one
brane as its ``left-hand'' boundary, $\del \mathcal{M}_L$, and the
other brane as its ``right-hand'' boundary, $\del \mathcal{M}_R$. The
entire bulk is now given by two copies of $\mathcal{M}$. 

Our set up is described by the following action
\be
S=S_\textrm{bulk}+S_\textrm{brane},
\ee
where the contribution from the bulk is given by
\be
S_\textrm{bulk}=2M^3 \int_\mathcal{M} \sqrt{g} (R-2\La) +\sum_{i=L, R}~
4M^3\int_{\del\mathcal{M}_i} \sqrt{g^{(i)}} K^{(i)}+2 \int_\mathcal{M}
L_m. 
\ee
Here $M$ is the five-dimensional Planck mass, and $g_{ab}$ is the bulk
metric with corresponding Ricci scalar, $R$. We have expilicitly included a
bulk cosmological constant, $\La$, which can be positive, negative or
zero. Any additional contributions from matter in the bulk appear in $L_m$. For $i=L$, $R$,
$g^{(i)}_{ab}$   is the induced metric on $\del\mathcal{M}_i$, and
$K^{(i)}_{ab}$ is the extrinsic curvature\footnote{We will adopt the
convention that $K^{(i)}_{ab}=\frac{1}{2}\mathcal{L}_n g^{(i)}_{ab}$,
the Lie derivative of the induced metric with respect to the {\it outward} pointing normal.} of  $\del\mathcal{M}_i$ in
$\mathcal{M}$.    Notice that $S_\textrm{bulk}$ contains an
overall factor of two because of the  $\mathbb{Z}_2$ symmetry.

The brane part of the action is given by
\be \label{eqn:braneaction}
S_\textrm{brane}=\sum_{i=L, R}~
\int_{\del\mathcal{M}_i}\sqrt{g^{(i)}}\left(-\sigma_i +2M^3 r_i\mathcal{R}^{(i)}\right)+\sum_{i=L, R}~
\int_{\del\mathcal{M}_i} \mathcal{L}_m^{(i)}
\ee
where for $i=L$, $R$, $\sigma_i$  is the tension of the brane at
$\del\mathcal{M}_i$. $\mathcal{L}_m^{(i)}$ gives the contribution from
any additional matter on the brane. We have also allowed for some localised curvature by
including the four-dimensional Ricci scalar,  $\mathcal{R}^{(i)}$ on
each brane. This is a DGP-like kinetic term~\cite{Dvali:DGPmodel},
whose contribution is weighted by the distance $r_i$ in each case. At
this stage we make no assumption about the magnitude, or even the sign
of the $r_i$.

The bulk equations of motion are given by the Einstein equations
\be \label{eqn:Einstein}
R_{ab}-\frac{1}{2} Rg_{ab}=-\La g_{ab}+\frac{1}{2M^3}T_{ab}
\ee
where $T_{ab}$ is the energy momentum tensor for additional matter in
the bulk. We also need to impose boundary conditions at the
branes. For $i=L,~R$, these are given by the $\mathbb{Z}_2$ symmetric Israel
equations~\cite{Israel:junction} at $\del \mathcal{M}_i$
\be \label{eqn:Israel}
K^{(i)}_{ab}=\frac{\sigma_{i}}{12M^3} g^{(i)}_{ab}-r_i\left
[ \mathcal{R}^{(i)}_{ab}-\frac{1}{6} \mathcal{R}^{(i)}
g^{(i)}_{ab}\right]+\frac{1}{2M^3} \left
[ \mathcal{T}^{(i)}_{ab}-\frac{1}{3}\mathcal{T}^{(i)} g^{(i)}_{ab}\right]
\ee
Here $ \mathcal{T}^{(i)}_{ab}$ corresponds to the energy momentum tensor
for additional matter on $\del \mathcal{M}_i$, and $\mathcal{T}^{(i)}$ is its trace.

\section{Background solutions} \label{sec:background}

In this section we will derive the metric, $\bar g_{ab}$, for the
background spacetimes. These correspond to solutions of the equations
of motion when no additional matter is present in the bulk, or on the
branes. In other words
\be
T_{ab}=\mathcal{T}^{(i)}_{ab}=0
\ee  
Let us introduce coordinates $x^a=(x^\mu, z)$, and  assume that
the brane at $\del\mathcal{M}_i$ satisfies $z=z_i$, where $z_L=0$ and
$z_R=l>0$. In order for us to trust our analysis, we need to
assume that
\be \label{eqn:quantumbound}
M > \sqrt{\frac{|\La|}{6}}, ~\frac{1}{l}
\ee
Now seek 
solutions of the form
\be
ds^2=\bar g_{ab}dx^adx^b=a^2(z) \tilde g_{\mu\nu}dx^\mu dx^\nu +dz^2.
\ee
where $\tilde g_{\mu\nu}$  describes a maximally
symmetric four-dimensional spacetime, with Riemann tensor
\be
\tilde R_{\mu \nu \al \beta}=\ka \la^2 \left(\tilde g_{\mu\al}\tilde g_{\nu\beta}-
\tilde g_{\mu\beta}\tilde g_{\nu\al} \right).
\ee 
where $\ka=0, \pm 1$ for Minkowski, de Sitter and anti-de Sitter space respectively.

The Einstein equations in the bulk give the following differential equations for $a(z)$
\be \label{eqn:a'}
\left(\frac{a^{\prime}}{a}\right)^2 = \ka \frac{
 \la^2}{a^2}-\frac{\La}{6}, \qquad
\frac{a^{\prime\prime}}{a}=-\frac{\La}{6}
\ee
By solving these, we find that for  $-l \leq z \leq
l$~\cite{Padilla:nested, Padilla:instantons, Padilla:thesis}
\be
a(z)=\begin{cases} \label{eqn:background}
\frac{\la}{k}\cos\left[\pm k(|z| -c) \right] &
\La=6k^2, \quad\ka=1 ~\textrm{only} \\
 \la\left[c \pm \ka |z|\right] & \La=0,\quad\quad\ka=0,~1 \\
\frac{\la}{2k}\left[e^{\pm k(|z|-c)}-\ka e^{\mp k(|z|-c)} \right] &
\La=-6k^2, ~\ka=0, \pm 1,
\end{cases}
\ee
where $c$ is an integration constant. We are free to define a length
scale by setting $a(0)=1$. This fixes a relationship between $\la$ and
$c$:
\be
\la=\begin{cases}
k/\cos(kc)&
\La=6k^2, \quad\ka=1 ~\textrm{only} \\
1/c  & \La=0,\quad\quad\ka=0,~1 \\
2k/\left[e^{\mp kc}-\ka e^{\pm kc} \right] &
\La=-6k^2, ~\ka=0, \pm 1,
\end{cases}
\ee
For each of the solutions (\ref{eqn:background}), the $\mathbb{Z}_2$ symmetry about $z=0$ is explicit,
whereas the symmetry about $z=l$ can be seen when we identify $z$
with $z+2l$. When we have a negative bulk cosmological constant, these
solutions correspond to those found in~\cite{Kaloper:bent,
Kim:inflation, Nihei:inflation,Karch:locally}. 

The Israel equations on $\del \mathcal{M}_i$ give us the following
boundary conditions
\be \label{eqn:bc}
\theta_i \frac{a^\prime (z_i)}{a(z_i)}= \frac{\sigma_{i}}{12M^3}-\ka r_i\frac{\la^2}{a^2(z_i)}
\ee
where $\theta_L=-1$ and  $\theta_R=+1$. For the special case of $\La
=-6k^2$, $\ka=0$,  these equations give the usual
Randall-Sundrum fine tuning of brane
tensions~\cite{Randall:hierarchy}. Finally, we note that
the cosmological constant on the brane at $\del \mathcal{M}_i$ is
given by
\be
\La_i=3\ka\frac{\la^2}{a^2(z_i)}.
\ee 
and shall henceforth refer to the branes as flat, de Sitter or anti-de
Sitter, when $\ka=0, \pm1$ respectively.  
\section{Metric Perturbations} \label{sec:pert}
We shall now consider perturbations about the background solutions~(\ref{eqn:background}) we
have just derived. We will allow additional matter to be present on
the branes, but not in the bulk. Let us define $g_{ab}=\bar
g_{ab}+\delta g_{ab}$ to be the perturbed
metric. We will work in Gaussian Normal coordinates so that
\be \label{eqn:GN}
\delta g_{\mu z}=\delta g_{zz}=0.
\ee
Since we have no additional bulk matter, we can take the metric to be
transverse-tracefree {\it in the bulk}. In other words, $\delta
g_{\mu\nu}=\chi_{\mu\nu}$, where
\be
\tilde \Del_\nu \chi^{\nu}_{\mu}=\chi^{\mu}_{\mu}=0.
\ee
Here $\tilde \Del$ is the covariant derivative with respect to the Minkowski,
de Sitter or anti de Sitter metric $\tilde g_{\mu\nu}$. In this choice
of gauge, the linearised bulk equations of
motion  are given by
\be \label{eqn:linbulk}
\left[ \frac{\tilde \Del^2-4\ka\la^2}{a^2}  + \frac{\del^2}{\del
z^2}+\frac{2\La}{3} \right] \chi_{\mu\nu}=0.
\ee
where we have used equation (\ref{eqn:a'}).

Unfortunately, we can no longer assume that the branes
are fixed at $z=z_i$. The presence of matter on the branes will cause them to bend~\cite{Garriga:gravity}. In general, they  will now be positioned at
$z=z_i+f_i(x)$, for some function $f_i$ that depends
only on the $x^{\mu}$. This makes it difficult to apply the Israel
equations at the branes. To get round this we apply a gauge
transformation that fixes $\del \mathcal{M}_L$, and another that fixes  $\del
\mathcal{M}_R$~\cite{Charmousis:radion}, without spoiling the Gaussian
Normal condition~(\ref{eqn:GN}). This gives rise to two
coordinate patches that are related by a gauge transformation in the
region of overlap. We will call the patch that fixes $\del
\mathcal{M}_L$,  the left-hand
coordinate patch, and the one that fixes $\del
\mathcal{M}_R$, the right-hand patch.

To fix  $\del \mathcal{M}_L$, we make the following coordinate transformation
\be
z \to z-f_L(x), \qquad x_\mu \to x_\mu +a^2(z)\del_\mu f_L
\int_{0}^{z} \frac{dy}{a^2(y)}
\ee
$\del \mathcal{M}_L$ is now fixed at $z=0$, although  $\del
\mathcal{M}_R$ is at $z=l+f_R-f_L$. The metric
perturbation in this patch is given by
\be
\delta g_{\mu\nu}=\chi^{(L)}_{\mu\nu}=\chi_{\mu\nu}-2a^2 \tilde
\Del_\mu \tilde \Del_\nu f_L
\int_{0}^{z} \frac{dy}{a^2(y)} +2\frac{a^\prime}{a}f_L\bar g_{\mu\nu}
\ee
Similarly, to fix  $\del \mathcal{M}_R$, let
\be
z \to z-f_R(x), \qquad  x_\mu \to x_\mu +a^2(z)\del_\mu f_R
\int_{l}^{z} \frac{dy}{a^2(y)}
\ee
Now we have  $\del \mathcal{M}_R$ at $z=l$ but with $\del
\mathcal{M}_L$ at $z=f_L-f_R$.  The metric perturbation is given by
\be
\delta g_{\mu\nu}=\chi^{(R)}_{\mu\nu}=\chi_{\mu\nu}-2a^2
\tilde \Del_\mu \tilde \Del_\nu f_R
\int_{l}^{z} \frac{dy}{a^2(y)} +2\frac{a^\prime}{a}f_R\bar g_{\mu\nu}
\ee
We are now ready to use the Israel equations (\ref{eqn:Israel}) to
give linearised boundary conditions at each brane. At $\del
\mathcal{M}_i$, we have
\begin{multline} \label{eqn:linbc}
\left[ -\frac{\theta_i}{2} \frac{\del}{\del
z}+\frac{\sigma_i}{12M^3}+\frac{r_i}{2a^2}\left( \tilde \Del^2-4 \ka
\la^2 \right) \right]\chi_{\mu\nu}=-\left(\theta_i+2r_i
\frac{a^{\prime}}{a} \right) \left[ \tilde \Del_\mu \tilde \Del_\nu
+\ka \frac{\la^2}{a^2} \bar{g}_{\mu\nu} \right]f_i \\-\frac{1}{2M^3}\left
[ \mathcal{T}^{(i)}_{\mu\nu}-\frac{1}{3}\mathcal{T}^{(i)} \bar g_{\mu\nu}\right]
\end{multline}
where everything is evaluated at $z=z_i$. In deriving equation
(\ref{eqn:linbc}), we have made repeated use of equations (\ref{eqn:a'})
and (\ref{eqn:bc}).

Note that if we take the trace of equation (\ref{eqn:linbc}), and use
the fact that $\chi_{\mu\nu}$ is transverse-tracefree, we get
\be \label{eqn:trace}
\left(\theta_i+2r_i
\frac{a^{\prime}(z_i)}{a(z_i)} \right)\left[ \frac{\tilde \Del^2+4\ka
\la^2}{a^2}\right] f_i=\frac{\mathcal{T}^{(i)}}{6M^3}
\ee
This clearly shows that matter on the brane does indeed cause the brane to
bend.

The general perturbation is made up of three parts: a {\it radion},
plane wave gravitons, and a particular solution that depends
explicitly on the matter sources, $\mathcal{T}_{\mu\nu}^{(i)}$. The
gravitons will be discussed in section~\ref{sec:spectrum}. In the next section we will concentrate on the radion.

\section{The Radion} \label{sec:radion}
\subsection{Radion wavefunction}
The radion mode, $\chi_{\mu\nu}^{\textrm{(rad)}}$, corresponds to fluctuations
in the brane separation. It is in addition to the fluctuation caused explicitly
by the matter on the brane, according to equation
(\ref{eqn:trace}). The radion exists because equation
(\ref{eqn:trace}) can have non-trivial solutions, $f_i$, even when
$\mathcal{T}^{(i)}_{\mu\nu}=0$. Clearly then, the best way to see the radion is to consider a situation where we have
no matter on the branes. Let us begin with a transverse-tracefree
gauge in the bulk
\be
\delta g_{\mu\nu}=\chi_{\mu\nu}^{\textrm{(rad)}}
\ee
with the branes positioned at $z=z_i+f_i(x)$. The fields, $f_i(x)$, are
yet to be determined. $\chi_{\mu\nu}^{\textrm{(rad)}}$ satisfies equation (\ref{eqn:linbulk}), with the following boundary
conditions at $\del \mathcal{M}_i$
\be \label{eqn:radbc}
\left[ -\frac{\theta_i}{2} \frac{\del}{\del
z}+\frac{\sigma_i}{12M^3}+\frac{r_i}{2a^2}\left( \tilde \Del^2-4 \ka
\la^2 \right) \right]\chi^{\textrm{(rad)}}_{\mu\nu}=-\left(\theta_i+2r_i
\frac{a^{\prime}}{a} \right) \left[ \tilde \Del_\mu \tilde \Del_\nu
+\ka \frac{\la^2}{a^2} \bar{g}_{\mu\nu} \right]f_i
\ee
where,  again, everything is evaluated at $z=z_i$. Motivated by the tensor
structure on the right-hand side of equation (\ref{eqn:linbc}), we look for
 solutions to (\ref{eqn:linbulk}) of the form
\be \label{eqn:rad}
\chi^{\textrm{(rad)}}_{\mu\nu}=\mu(z)\left( \tilde \Del_\mu\tilde \Del_\nu+\ka
\frac{\la^2}{a^2}\bar g_{\mu\nu} \right) \phi
\ee
where we have introduced a scalar field, $\phi(x)$, which we will call
the radion field. Note that since $\chi^{\textrm{(rad)}}_{\mu\nu}$ is transverse-tracefree, we must have
 \be
\left[ \tilde \Del^2+4\ka
\la^2\right] \phi=0
\ee
This means that the radion field, $\phi$,
has mass $m^2=-4\ka
\la^2$. When $\ka=0$, the radion is massless, whereas for $\ka=-1$, it
is massive. For $\ka=1$, the radion is tachyonic, signalling an
instability for de Sitter branes~\cite{Chacko:radion}.

Given the ansatz (\ref{eqn:rad}), we solve (\ref{eqn:linbulk}) to get
an expression for $\mu(z)$.
\be
\mu(z)=\begin{cases}
\al+\beta z & \textrm{for $\La=\ka=0$} \\
aa^{\prime}\left[\al
+\beta \int^z\frac{dy}{\left(a a^{\prime}\right)^2} \right] &
\textrm{otherwise}
\end{cases} 
\ee
However, these expressions for $\mu(z)$ are too general. We find that $\chi_{\mu\nu}^{\textrm{(rad)}}$ contains terms which are
pure gauge, and therefore unphysical. In fact, when $\La=\ka=0$,
$\chi_{\mu\nu}^{\textrm{(rad)}}$ can be completely gauged away by the
following coordinate tranformation
\be
x^{\mu} \to x^{\mu}+(\al+\beta z)\del_\mu \phi, \qquad z \to z-\beta
\phi
\ee
When we apply the boundary conditions (\ref{eqn:linbc}), we find that
$f_L=f_R=0$. This means that there is no physical radion, and no matter-independant brane bending.

For every other case, the radion cannot be gauged away completely, so
we are left with a physical radion mode. We {\it can} remove the term
proportional to $\al$ in  $\chi_{\mu\nu}^{\textrm{(rad)}}$ by taking
\be
x^{\mu} \to x^{\mu}+\frac{1}{2} \al aa^{\prime} \del_\mu \phi, \qquad  z
\to z+\frac{1}{2}\al \ka \la^2 \phi.
\ee
but we can do nothing about the other term. The {\it physical} radion perturbation is
therefore given by
equation (\ref{eqn:rad}) with
\be
\mu(z) =
 a a^{\prime}\int^z\frac{dy}{\left(a a^{\prime}\right)^2} 
\ee
Notice that we have  set $\beta=1$ by absorbing it into the definition of
$\phi$. 
Finally, the boundary conditions (\ref{eqn:radbc}) give the $f_i$ in
terms of $\phi(x)$. Making use of equations (\ref{eqn:a'}) and (\ref{eqn:bc}), we get
\be \label{eqn:phi}
f_i(x)=\frac{1}{2}\left[\frac{1~}{v_ia(z_i)a^\prime(z_i)}-\ka \la^2\int^{z_i}\frac{dy}{\left(a a^{\prime}\right)^2}\right]\phi(x)
\ee
where 
\be
v_i=1+2\theta_i r_i \frac{a^{\prime}(z_i)}{a(z_i)}
\ee
Note that equation (\ref{eqn:phi}) also imposes a relationship between
$f_L(x)$ and $f_R(x)$.

\subsection{Radion effective action}
In braneworld theories that exhibit modification of gravity at
large distances, we often find that the radion field is a ghost (see,
for example,~\cite{Gregory:GRS, Pilo:ghost}). We therefore need to
check whether this is the case in any of our models. We do this by
calculating the radion effective action to quadratic order,
following~\cite{Pilo:ghost} (see~\cite{Bagger:radion} for the
supersymmetric case).

Let us work with the left-hand coordinate patch, which has the branes positioned at
$z=0$ and $z=l+\delta l(x)$, where $\delta
l=f_R-f_L$. The metric perturbation is given by
\be
\delta g_{\mu\nu}=\chi_{\mu\nu}^{\textrm{(rad)}}-2a^2 \tilde
\Del_\mu \tilde \Del_\nu f_L
\int_{0}^{z} \frac{dy}{a^2(y)} +2\frac{a^\prime}{a}f_L\bar g_{\mu\nu},
\qquad \delta  g_{\mu z}= \delta  g_{zz}=0
\ee
In order to integrate out the extra
dimension, it is convenient to have {\it both} branes fixed. We can do
this with the following coordinate transformation
\be
z \to z- B(z)\delta l ,\qquad  x_{\mu} \to x_{\mu}+a^2(z)\del_{\mu} (\delta
l)\int^z_0 \frac{B(y)}{a^2(y)} dy
\ee
where $B(z)$ is some differentiable function for $0 \leq z \leq l$,
satisfying $B(0)=0$ and $B(l)=1$. While this transformation ensures
that $\delta g_{\mu z}$ is still zero, the price we pay for fixed branes
is that we now have non-vanishing $\delta g_{zz}$. More precisely,
\begin{eqnarray}
&&\delta g_{\mu\nu}=h_{\mu\nu} = \chi_{\mu\nu}^\textrm{(rad)}+2\frac{a^\prime}{a}\left[f_L+\delta l B(z)\right]\bar g_{\mu\nu}-2a^2 \tilde
\Del_\mu \tilde \Del_\nu f_L
\int_{0}^{z} \frac{dy}{a^2(y)}\nonumber
\\&& \qquad\qquad\qquad\qquad\qquad\qquad  -2a^2 \tilde
\Del_\mu\Del_\nu(\delta l)\int_0^z \frac{B(y)}{a^2(y)} dy\\
&&\delta g_{zz}=h_{zz} =2\delta lB^\prime(z) \label{fixed}
\end{eqnarray} 
To quadratic order, the effective action is given by 
\be \label{eqn:effaction}
S_\textrm{eff}=-M^3 \int_0^l dz \int d^4x \sqrt{\bar g} ~
h^{ab}\delta E_{ab}-\sum_{i=L, R}M^3\int_{z=z_i} d^4x \sqrt{\bar g} ~
h^{\mu\nu}\delta\Theta^{(i)}_{\mu\nu}
\ee
where $\delta E_{ab}$ and $\delta \Theta^{(i)}_{\mu\nu}$ are
the expansions, to linear order, of the bulk and boundary
equations of motion respectively.
\begin{eqnarray}
E_{ab} &=& R_{ab}-\frac{1}{2}Rg_{ab}+\La g_{ab}
\\
\Theta^{(i)}_{\mu\nu} &=& K^{(i)}_{\mu\nu}- K^{(i)}g^{(i)}_{\mu\nu}+\frac{\sigma_i}{4M^3}g^{(i)}_{\mu\nu}+r_i\left[\mathcal{R}^{(i)}_{\mu\nu}-\frac{1}{2}\mathcal{R}^{(i)}g^{(i)}_{\mu\nu}\right]
\end{eqnarray}
We find that $\delta E_{\mu\nu}$ and $\delta \Theta^{(i)}_{\mu\nu}$
are identically zero~\cite{Pilo:ghost}. After integrating out a
total derivative in $z$, we arrive at the following effective action
for the radion
\be
S_\textrm{eff}=-3M^3 \left[ \Delta
\left(\frac{1}{v_ia(z_i)a^\prime(z_i)}\right)-\ka \la^2 \int_0^l
\frac{dy}{\left(aa^\prime\right)^2} \right]\int d^4x \sqrt{\tilde g}~\phi\left( \tilde \Del^2+4
\ka \la^2 \right)\phi
\ee
where $\Delta Q_i=Q_R-Q_L$.

\subsection{``No ghost'' bounds} \label{subsec:radghost?}
The radion field, $\phi$, becomes a ghost when the coefficient of its
kinetic term becomes negative. This would indicate a sickness in our
theory. We therefore require that
\be \label{eqn:ghost?}
 \Delta
\left(\frac{1}{v_ia(z_i)a^\prime(z_i)}\right)-\ka \la^2 \int_0^l
\frac{dy}{\left(aa^\prime\right)^2} \leqslant 0
\ee
We shall not consider what this means for de Sitter branes, because
the radion is tachyonic, even if it is not a ghost. However, for flat
and anti-de Sitter branes it is worth considering this condition in
more detail.

When $\La=\ka=0$, there is no radion, and therefore nothing to worry
about. When $\La=-6k^2$,  $\ka=0$, we have $a(z)=e^{-k|z|}$, and the
condition (\ref{eqn:ghost?}) demands that
\be \label{RSghostrad?}
\frac{1}{1-2kr_R} \geqslant \frac{e^{-2kl}}{1+2kr_L}
\ee
Now consider anti-de Sitter branes with $\La=-6k^2$, $\ka=-1$. Given
that the
warp factor
$a(z)=\frac{\la}{k}\cosh\left[k\left(|z|-c\right)\right]$, we can
show that the
condition (\ref{eqn:ghost?}) requires that
\be \label{eqn:adsghost?}
\frac{\frac{\sinh(2kc)}{1+2kr_R\tanh[k(l-c)]}+\frac{\sinh[2k(l-c)]}{1+2kr_L\tanh(kc)}-\sinh(2kl)}{\sinh[2k(l-c)]\sinh(2kc)}\leqslant 0
\ee
Here we will mainly be interested in the ``symmetric'' case ($c=l/2$),
so that (\ref{eqn:adsghost?}) simplifies to
\be \label{adsghostrad?}
\frac{1}{1+2kr_R\tanh(kl/2)}+\frac{1}{1+2kr_L\tanh(kl/2)} \leqslant 2\cosh(kl)
\ee
To sum up, we have shown that we can avoid  ghost-like radions by
placing certain bounds on the parameters in our model. While this is
of no real use for tachyonic de Sitter branes, it will be important when considering flat and anti-de Sitter branes.

\section{Gravitons} \label{sec:spectrum}
In this section we will consider graviton perturbations in detail. We
will ask two important questions: (i) when does the graviton become a
ghost, and (ii) do we have a {\it bigravity} model? As with the
radion, we can derive ``no ghost'' bounds for the graviton by
calculating the effective action. To establish whether or not we have
bigravity, we need to find the graviton mass spectrum. For bigravity,
we expect to see a massless mode and an {\it ultralight} mode~\cite{Kogan:2000, Kogan:branemulti, Kogan:modification,
Kogan:adsbranes, Kogan:6dmulti, Kogan:review, Papazoglou:multithesis,
Damour:univclass, Damour:nonlin}.  This will lead to modifications of
gravity at large distances. However, it is important to note
that this modified behaviour may never be observable. For example, for AdS branes there is a danger that the Compton wavelength of the
ultralight graviton is larger than the horizon
size~\cite{Kogan:adsbranes}. Once we have established the
existance of an 
ultralight graviton, we will check that it could one day lead to
interesting observations.
\subsection{Graviton mass spectrum} \label{subsec:spectrum}
Consider the metric perturbation decomposed into plane waves
\be \label{eqn:pw}
\delta g_{\mu\nu}(x,z)=\sum_{m} \chi_{\mu\nu}^{(m)}(x)u_m(z)
\ee
where 
\be
\left(\tilde \Del^2-2\ka \la^2\right)\chi_{\mu\nu}^{(m)}=m^2 \chi_{\mu\nu}^{(m)}.
\ee
From a four-dimensional perspective, $\chi_{\mu\nu}^{(m)}$
corresponds to a transverse-tracefree, spin 2 particle of mass,
$m$~\cite{Gibbons:instanton,Allen:vector2pt,
Allen:dsprop1,Turyn:curvfluc, Turyn:maxsymprop, Allen:dsprop2,
Chodos:casimir, Higuchi:dsprop, D'Hoker:adsprop, Naqvi:adsprop}. 

We now insert (\ref{eqn:pw})
into the equations of motion (\ref{eqn:linbulk}) and (\ref{eqn:linbc}). Since we are interested in
plane waves, and the radion has already been accounted for, we set
$f_i=\mathcal{T}^{(i)}_{\mu\nu}=0$. The equations of motion become
\be \label{eqn:specbulk}
\left(\frac{m^2-2\ka\la^2}{a^2}  + \frac{\del^2}{\del
z^2}+ \frac{2\La}{3}\right)u_m(z)=0, 
\ee
in the bulk, with boundary conditions
\be \label{eqn:specbc}
\left[ -\frac{\theta_i}{2}\frac{ \del}{\partial z}+\frac{\sigma_i}{12M^3}+\frac{r_i}{2a^2}\left( m^2-2\ka
\la^2 \right) \right]u_m(z)=0
\ee
at $z=z_i$.  Different eigenstates are orthogonal, so that
for  $m \neq n$, 
\be \label{eqn:normalisation}
\int_{-l}^{l} dz~\frac{1}{a^{2}(z)} \left[1+2r_L \delta
(z)+2r_R \delta
(z-l) \right] u_m(z)u_n(z)=0
\ee
We can always find a zero mode that satisfies the boundary
conditions
\be \label{eqn:zero}
u_0(z)=A_0a^2(z)
\ee
where $A_0$ is some normalisation constant. We
shall now look for massive modes for the flat and anti-de Sitter
branes of primary interest, taking each case in turn.

\subsubsection{$\La=\ka=0$} \label{sec:La=ka=0}
Since $a(z) \equiv 1$, the bulk equation (\ref{eqn:specbulk}) is
particularly simple. In addition to the zero mode (\ref{eqn:zero}), we find a number of massive modes
\be \label{eqn:flatwvfn1}
u_m(z)=A_m\cos(m|z|)+B_m\sin(m|z|)
\ee
for some constants $A_m$, $B_m$, that are yet to be determined. The
boundary conditions require that 
\begin{equation}
r_LmA_m+B_m \label{eqn:flatwvfn2}
=0
\ee
\begin{equation} 
\left[r_Rm\cos(ml)+\sin(ml)
\right]A_m+\left[r_Rm\sin(ml)-\cos(ml)
\right]B_m =0
\end{equation} 
Note that we have used the fact that $\sigma_i=0$. If we want a
non-trivial solution for $A_m$ and $B_m$, we are only allowed
quantised values of $m$, satisfying
\be \label{eqn:qum}
\left(m^2r_Lr_R-1\right)\sin(ml)=m\left(r_L+r_R\right)\cos(ml). 
\ee
The main KK tower will therefore be made up of states with $m_\textrm{heavy}\gtrsim  1/l$. Are there any ultralight states? To answer
this, consider (\ref{eqn:qum}) when $m \ll 1/l$. We find a mode with mass
\be \label{eqn:flatlightmass}
m_\textrm{light} \approx \sqrt{\frac{6\left(l+r_L+r_R\right)}{l^3+3l^2(r_L+r_R)+6lr_Lr_R}}
\ee
This is indeed much smaller than $1/l$, if we choose $x=r_L/l$ and
$y=r_R/l$ so that
\be \label{bigrav:flat}
\epsilon^2=\frac{\left(1+x+y\right)}{1+3(x+y)+6xy}
\ee
where $| \epsilon | \ll 1$.

\subsubsection{$\La=-6k^2$; $\ka=0$}
We now consider the Randall-Sundrum I model with brane-localised
curvature terms. This time we have $a(z)=e^{-k|z|}$. By solving the
bulk equation (\ref{eqn:specbulk}), we again find a number of massive modes~\cite{Randall:alt, Padilla:thesis}
\be
u_m(z)=A_mJ_2(me^{k|z|}/k)+B_mY_2(me^{k|z|}/k)
\ee
where $J_n$ and $Y_n$ are Bessel's functions of integer order $n$. The
boundary conditions (\ref{eqn:specbc}) imply 
\begin{multline} \label{eqn:RSbc}
A_m\left[-\theta_i J_1(me^{kz_i}/k)+mr_i
e^{kz_i}J_2(me^{kz_i}/k)\right]
\\+B_m\left[-\theta_i Y_1(me^{kz_i}/k)+mr_i e^{kz_i}Y_2(me^{kz_i}/k)\right]=0
\end{multline}
On this occasion we have used the fact
$-k\theta_i=\sigma_i/12M^3$.  Equation (\ref{eqn:RSbc}) is of course
{\it two} equations, one for $i=L$ and one for $i=R$. For these to
have non-trivial solutions for $A_m$ and $B_m$, we again find that $m$
is quantised accordingly
\begin{multline} \label{eqn:RSqu}
\left[J_1(m/k)+mr_LJ_2(m/k)\right]\left[-Y_1(me^{kl}/k)+mr_R
e^{kl}Y_2(me^{kl}/k)\right] \\=\left[Y_1(m/k)+mr_LY_2(m/k)\right]\left[-J_1(me^{kl}/k)+mr_R e^{kl}J_2(me^{kl}/k)\right]
\end{multline}
We now focus on the case when the brane separation is large, $l \gtrsim
1/k$. In this instance, the main KK tower is made up of states with
mass $m_\textrm{heavy} \gtrsim k e^{-kl}$~\cite{Randall:alt,
Padilla:thesis,Davoudiasl:braneloc}. Is there an ultralight state?
Consider (\ref{eqn:RSqu}) when $m \ll k e^{-kl}$, and use the fact
that
\begin{align}
J_1(t)&=\frac{t}{2}-\frac{t^3}{16}+\mathcal{O}\left(t^{5}\right) 
&Y_1(t)&=-\frac{2}{\pi t}+\frac{t}{\pi}\ln\left(\frac{t}{2} \right)+\mathcal{O}\left(t\right) \\
J_2(t)&=\frac{t^2}{8} +\mathcal{O}\left(t^{4}\right)  &Y_2(t)&=-\frac{4}{\pi t^2}-\frac{1}{\pi}\mathcal{O}\left(t^2\ln(t)\right)
\end{align}
for $|t| \ll 1$. To leading order, we find a mode with mass
\be \label{eqn:RSlightmass}
m_\textrm{light} \approx 2ke^{-kl}\sqrt{\frac{\left(kr_L+\frac{1}{2}\right)e^{2kl}+kr_R-\frac{1}{2}}{\left(kr_L+\frac{1}{2}\right)\left(kr_R+\frac{1}{2}\right)e^{2kl}-k(l+r_L+r_R)-\left(kr_L-\frac{1}{2}\right)\left(kr_R-\frac{1}{2}\right)e^{-2kl}}}
\ee
This is much smaller than $ke^{-kl}$, as long as we choose $x=kr_L$
and $y=kr_R$, so that
\be \label{bigrav:RS}
\epsilon^2=\frac{\left(x+\frac{1}{2}\right)e^{2kl}+y-\frac{1}{2}}{\left(x+\frac{1}{2}\right)\left(y+\frac{1}{2}\right)e^{2kl}-(kl+x+y)-\left(x-\frac{1}{2}\right)\left(y-\frac{1}{2}\right)e^{-2kl}}
\ee
where $|\epsilon| \ll 1$. 

Now consider the opposite regime, when the brane separation is small,
$l \ll 1/k$. The mass spectrum is well approximated by the expressions
given in section~\ref{sec:La=ka=0}, when $\La=\ka=0$: the heavy modes
have mass $m_\textrm{heavy} \gtrsim 1/l$, and the mass of the light mode is given by
(\ref{eqn:flatlightmass}). Note that
(\ref{eqn:RSlightmass}) reduces to (\ref{eqn:flatlightmass}) in this
limit, and is therefore valid for all values of $l$.

\subsubsection{$\La=-6k^2$; $\ka=-1$}
Finally, we consider AdS branes with brane-localised curvature. We shall focus on the
``symmetric'' case, where
$a(z)=\frac{\la}{k}\cosh\left[k\left(|z|-\frac{l}{2}\right)\right]$
and $\la=k/\cosh\left[kl/2\right]$. A detailed analysis can be found
in appendix~\ref{app:spectrum}. There we see that the eigenstates, $u_m(z)$, are
given in terms of hypergeometric functions (see equation
(\ref{eqn:adsum})). Again, the  boundary conditions impose a
quantisation condition on $m$ (see equation
(\ref{eqn:adsqu})).

When the branes are far apart ($l \gtrsim 1/k$), the situation is very similar to the case when $r_i=0$~\cite{Kogan:adsbranes}. We have
bigravity, but the Compton wavelength of the ultralight graviton is
larger than the size of the AdS horizon. Large distance modifications
of gravity will therefore never be observable.

Now consider the opposite regime, when the branes are very close
together ($l \ll 1/k$). Again, the expressions given in
section~\ref{sec:La=ka=0} give a good estimate for the mass
spectrum. The horizon size on either brane is given by $1/\lambda
\approx 1/k$. Large distance modifications of gravity could therefore be
observable, as long as, 
\be
\frac{1}{l} \gg m_\textrm{light} >  k
\ee

\subsection{Graviton effective action}
Now that we know the graviton mass spectrum, we are ready to calculate
the corresponding effective action to check for ghosts. Since we have assumed that
$f_i=0$, we can immediately insert the metric (\ref{eqn:pw}) into
equation (\ref{eqn:effaction}). Making use of the orthogonality
condition (\ref{eqn:normalisation}), we find that the effective
action, to quadratic order, is given by
\be
S_\textrm{eff}=\sum_{m}C_m \int d^4x \sqrt{\tilde g} ~\chi^{(m) \mu\nu} \left(\tilde
\nabla^2-2\ka\la^2-m^2\right) \chi^{(m)}_{\mu\nu}
\ee
where
\be
C_m=\frac{M^3}{2}\int_{0}^{l}dz~
\left[\frac{u_m(z)}{a(z)}\right]^2\left[1+r_L\delta(z)+r_R\delta(z-l)\right]
\ee
As an aside, we will also examine the coupling of each mode to matter
on the branes.
To do this, we allow for
$\mathcal{T}^{(i)}_{\mu\nu} \neq 0$, giving additional terms
\be
S_\textrm{matter}=\sum_{i=L, R} \sum_{m} \int d^4x \sqrt{\tilde g} ~\frac{u_m(z_i)}{2}
\chi^{(m) \mu\nu} \mathcal{T}_{\mu\nu}^{(i)}
\ee
If we define the canonically normalised fields as
\be
\widehat \chi^{(m)}_{\mu\nu}=\sqrt{C_m}\chi^{(m)}_{\mu\nu}
\ee
we see that the mode of mass, $m$  has the following coupling to
matter on $\partial \mathcal{M}_i$,
\be
g^{(i)}_m=\frac{u_m(z_i)}{2\sqrt{C_m}}.
\ee
We can relate these couplings to the four-dimensional Planck mass,
$M_i$ on  $\partial \mathcal{M}_i$.  At the intermediate energy scale, $m_\textrm{heavy} \gg p
\gg m_\textrm{light}$, gravity is due to both the
massless and ultralight mode.  We deduce that,
\be
\frac{1}{M_i^2}=2\left[g^{(i)}_0\right]^2+2\left[g^{(i)}_\textrm{light}\right]^2
\ee
\subsection{``No ghost'' bounds} \label{subsec:gravghost?}
We can now ask whether or not we have any ghost-like gravitons. If we assume that $\chi^{(m)}_{\mu\nu}$ is real, it is clear that this
mode is {\it not} a ghost, provided
\be
C_m \geqslant 0
\ee
Since we are not too concerned about the ultra-violet behaviour of
this theory, we only need to impose this condition on the massless and
ultralight modes. For the massless mode (\ref{eqn:zero}), this is equivalent to
\be \label{condition:massless}
\int_0^l a^2(z)~dz +\sum_{i=L, R} r_i a^2(z_i) \geqslant 0
\ee
For each case of interest, this gives
\ba
l+r_L+r_R \geqslant 0 &&\textrm{for} ~\La=\ka=0 \label{propflat}\\
\left(kr_L+\frac{1}{2}\right)e^{2kl} + kr_R -\frac{1}{2} \geqslant0
&&\textrm{for}~\La=-6k^2, ~\ka=0  \label{propRS}\\
k\left(r_L+r_R\right)
+\tanh\left(\frac{kl}{2}\right)+\frac{kl}{2}\sech^2\left(\frac{kl}{2}\right)\geqslant
0
&&\textrm{for ``symmetric'' AdS branes} \label{propAdS} \qquad
\ea
In addition we also require that $C_\textrm{light} \geqslant
0$. However, it will not be necessary to evaluate this explicitly
in all cases, as most bigravity models
will be ruled out by the radion bounds (\ref{eqn:ghost?}) and the
massless mode condition (\ref{condition:massless}).

\section{Analysis} \label{sec:analysis}
The aim of this paper is to find a braneworld bigravity theory that is
free of ghosts and tachyons. We saw in section~\ref{sec:radion} that
tachyons appear on de Sitter branes in the form of the radion field,
so we ruled these models out. Other 
types of brane have not been ruled out, and we showed in
section~\ref{subsec:spectrum} how one can choose parameters so that we get
a model of bigravity. In this section we will check whether or not
these choices are compatible with the ``no ghost'' bounds of
sections~\ref{subsec:radghost?} and~\ref{subsec:gravghost?}. We will also keep in mind the validity bound
(\ref{eqn:quantumbound}), and the hope that bigravity will lead to
modifications of gravity that will one day be observable.

\subsection{$\La=\ka=0$}

Let us begin with the case $\La=\ka=0$. For bigravity, we need
equation (\ref{bigrav:flat}) to hold for $|\epsilon| \ll 1$ (recall that
$x=r_L/l$ and $y=r_R/l$). We have no radion to worry about, although
we do need to be careful with the gravitons. The massless
graviton is {\it not} a ghost, as long as
\be \label{bound:flat}
1+x+y \geqslant 0
\ee
Now consider the plot of $x=r_L/l$ against $y=r_R/l$ shown in
figure~\ref{fig:flat}. 
\begin{figure}
\begin{center}
\includegraphics[width=10cm, height=10cm]{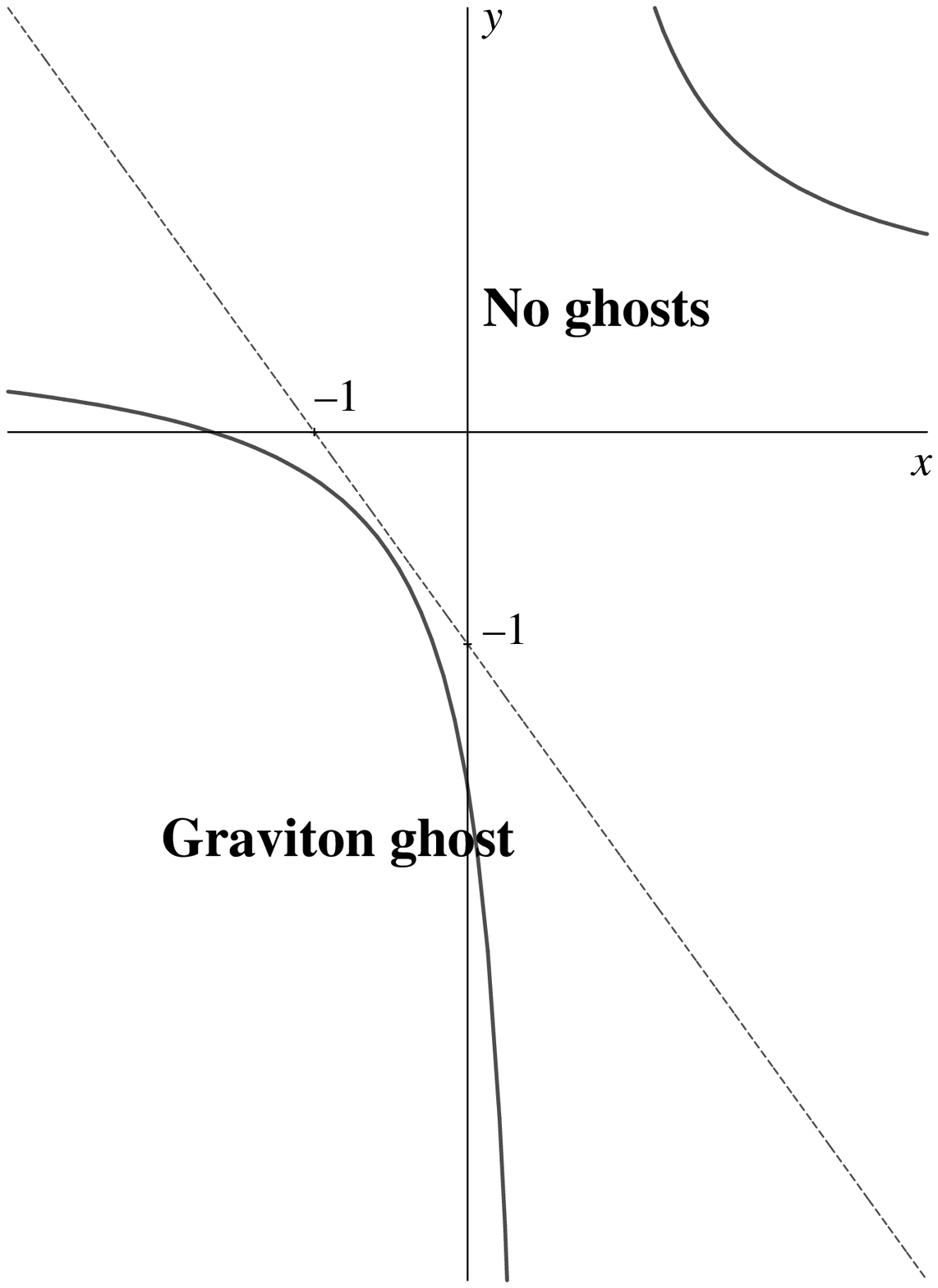}
\vskip 5mm
\caption{Analysis for $\La=\ka=0$.} \label{fig:flat}
\end{center}
\end{figure}
The solid black line corresponds to a line of
constant $|\epsilon| \ll 1$. As $x \to \infty$, $y \to 1/6\epsilon^2$,
and vice-versa. The dotted line corresponds to $1+x+y=0$, and
represents the boundary between the massless graviton being well behaved, and the appearance of a ghost. Since the solid line never crosses the dotted
line, it is clear that we need $x, ~y \geqslant 1/6\epsilon^2$. In other words,
we need $r_i \gg l$.

We are now ready to check the behaviour of the ultralight mode. For simplicity, let $r_L=r_R =r
\gg l$, so that 
\be
m_\textrm{light} \approx \sqrt{\frac{2~}{lr}}
\ee
If we use this along with equations (\ref{eqn:flatwvfn1}) and (\ref{eqn:flatwvfn2}), we find that
\be
C_\textrm{light}\approx A_m^2M^3 r \geqslant 0
\ee
The ultralight mode is not ghost, so we conclude that we have indeed
found a bigravity model that is ghost-free. At intermediate energy
scales, note that the Planck mass on either brane is given by
\be \label{Planck}
M_i^2=m^2_{pl} \approx M^3 r
\ee
as we might have expected.

\subsection{$\La=-6k^2;  ~l \ll 1/k;~\ka=0,-1$}
We now turn our attention to the case where $\La=-6k^2, ~\ka=0,-1$,
with $l \ll 1/k$. In this limit, the mass spectrum is the same as for
$\La=\ka=0$, and the graviton bounds (\ref{propRS}) and
(\ref{propAdS}), are reduced to (\ref{bound:flat}). The analysis will
therefore be very similar to what we have just done for $\La=\ka=0$,
as we would expect by continuity. There are, however, a couple of other
things to consider now. We have a radion to worry about (see equations (\ref{RSghostrad?}) and
(\ref{adsghostrad?})), and for $\ka=-1$, we need to ensure that any
infra-red modifications of gravity will not be hidden behind the AdS horizon.

For simplicity, let us again assume that $r_L=r_R=r \gg l$. In both
cases we find that there is an upper bound on the value of $r$
\be
r < \begin{cases} 1/2k & \textrm{for $\ka=0$} \\
2/k^2l &  \textrm{for $\ka=-1$}
\end{cases}
\ee
For $\ka=0$, this is due to the radion becoming a
ghost~\cite{Davoudiasl:braneloc}. For $\ka=-1$, however, the radion is
not a problem, and the bound is due to the condition,
$m_\textrm{light} >k$. This ensures that long distance modifiactions
of gravity may one day be observable.

The upper bounds on $r$ are  not too troublesome, as long as we take $k$ to be
very small. 

\subsection{$\La=-6k^2;  ~l \gtrsim 1/k;~\ka=0$}
Finally, we consider the (DGP) extension of the RS model, with the branes
far apart. For bigravity, we need
equation (\ref{bigrav:RS}) to hold for $|\epsilon| \ll 1$ (recall that
now we have $x=kr_L$ and $y=kr_R$). From equations (\ref{RSghostrad?}) and
(\ref{propRS}) we have the following bounds
\ba
\frac{-1~}{y-\frac{1}{2}} \geqslant \frac{e^{-2kl}}{x+\frac{1}{2}} && \textrm{to avoid a
radion ghost} \\
x+\frac{1}{2}+\left(y-\frac{1}{2}\right)e^{-2kl} \geqslant 0 &&\textrm{to avoid
a (massless) graviton ghost}
\ea
\begin{figure}
\begin{center}
\includegraphics[width=10cm, height=10cm]{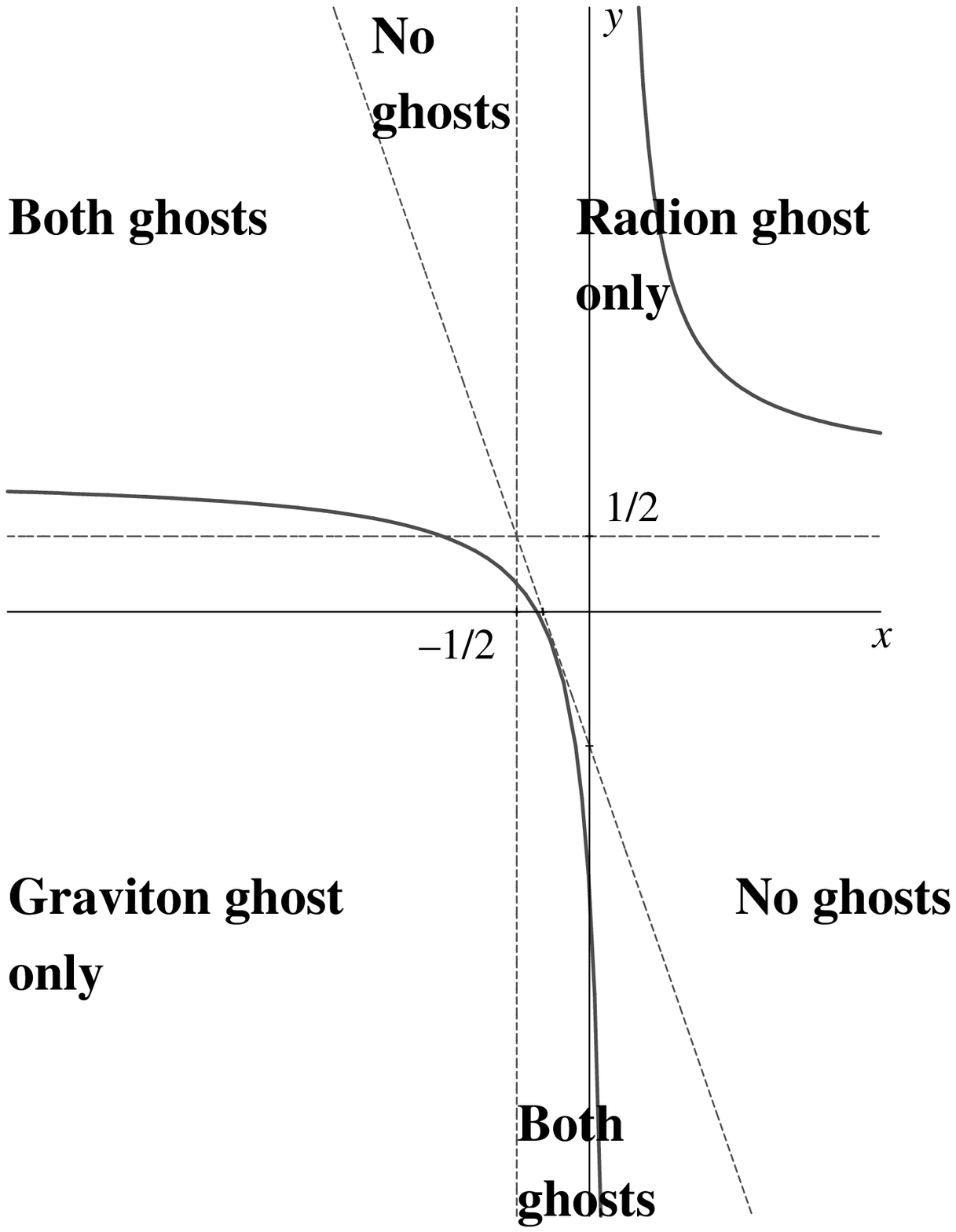}
\vskip 5mm
\caption{Analysis for $\La=-6k^2$, ~$\ka=0$, ~$l\gtrsim 1/k$.} \label{fig:RS}
\end{center}
\end{figure}
Now consider the plot of $x=kr_L$ against $y=kr_R$ shown in
figure~\ref{fig:RS}. Again, the solid black line corresponds to a
line of constant $|\epsilon| \ll 1$. The asymptotic behaviour is
\ba
y \to
\frac{1+\coth(2kl)}{2\epsilon^2} &&
~\textrm{as}~x \to \infty \nonumber\\
x \to
\frac{\cosech(2kl)}{2\epsilon^2} &&
~\textrm{as}~y \to \infty \nonumber
\ea
The dotted lines correspond to boundaries, across which the
``ghost-like'' status of the theory changes. The diagonal line is
given by $x+\frac{1}{2}+\left(y-\frac{1}{2}\right)e^{-2kl}=0$, whereas
the horizontal and vertical lines are given by $y=1/2$ and $x=-1/2$
respectively. It is easy to check that the solid line never enters any
of the ``no ghost'' regions indicated. This means that for this model
to exhibit bigravity, either the radion, the graviton, or both,
will become a ghost. The model is therefore rejected.
\section{Conclusions} \label{sec:conclusions}
In this paper we have studied a class of braneworld models, with
localised curvature on the branes. Some of these models give rise to
bigravity, leading to large distance modifications of gravity for a
four-dimensional observer. After checking for ghosts and tachyons, we
rejected most, but not all of the models we had considered.

One of these well behaved models consists of a flat bulk ($\La=0$)
sandwiched in between two flat branes ($\ka=0$) of zero tension. The
branes are close together, and the curvature terms on the brane are
large and positive ($r_i \gg l$). The mass spectrum is precisely that of
bigravity. We have a massless graviton, an ultralight graviton, and a
tower of heavy KK modes. The model is completely free of ghosts and
tachyons.

To add some substance, let us put some numbers in. If we take
$l \sim 10^3 ~(\textrm{eV})^{-1}$, and $r_i \sim 10^{59} ~(\textrm{eV})^{-1}$, we find that
\be
m_\textrm{light} \sim 10^{-31} ~\textrm{eV}, \qquad m_\textrm{heavy} \gtrsim
10^{-3} ~\textrm{eV}.
\ee
These masses lie outside of the range for which gravity is well
tested (\ref{range}), so we have no contradiction with experiment. Since the
four-dimensional Planck mass, $m_{pl} \sim 10^{18}~\textrm{GeV}$, we
conclude from equation (\ref{Planck}) that $M \sim
10^{-2}-10^{-1}~\textrm{eV}$. While the fundamental Planck scale is
very low, it does not violate the validity
bound (\ref{eqn:quantumbound}), and exceeds the scale probed by
Cavendish experiments ($10^{-3}~\textrm{eV}$). In principle
there are cosmological and astrophysical bounds that one should
consider, such as the effect on star cooling due to graviton emmision
into the bulk. This will be left for
future research.

Note that $m_\textrm{light} \sim 10^{-31} ~\textrm{eV}$ is
just about heavy enough for us to expect interesting observations
today. This could be important in trying to explain cosmic
acceleration. If we wanted to reduce this mass even
further, we would have to increase $r$. Even a small increase would
push the Planck mass below the Cavendish scale. Curiously, we seem to
be in just the right place to start seeing  interesting new physics,
either in the infra-red, or the ultra-violet\footnote{I would like to
thank John March-Russell for this observation}. 

We can go beyond this model by switching on a small AdS curvature in
the bulk, and still get
ghost-free bigravity. Although a tachyon appears for de Sitter
branes, this is not the case for flat and anti-de Sitter branes. When
the bulk AdS length is of the same order as the brane separation, we
find that ghost-free bigravity becomes impossible. 

By checking for ghosts and tachyons in our models, we have carried out
the first
important tests of viability. However, we should be aware that there
are other possible problems. As with all models of massive gravity, we should be concerned with the
famous vDVZ discontinuity~\cite{vanDam:VDVZ, Zakharov:VDVZ}. Our
linearised equations of motion (\ref{eqn:linbc}) suggest that this may
well be an issue. This is especially true for the case $\La=\ka=0$,
as there is no ``brane-bending'' effect that could cancel off any
unwanted degrees of freedom (see, for example,
~\cite{Giddings:lingrav, Csaki:props}).  Even our AdS brane model will
suffer this problem. This is a surprise, as it is often said that there is no vDVZ discontinuity
in AdS space~\cite{Kogan:VDVZ, Porrati:VDVZ}. However, this result relies on the
graviton mass going to zero faster than the inverse horizon size. In
order to ensure that we had {\it observable} bigravity, we demanded
the opposite of this.

Crucially, there may be a way around this problem. Vainshtein {\it et al}~\cite{Vainshtein:nonVDVZ,Deffayet:noVDVZ}, have argued that the perturbative expansion in
Newton's constant in~\cite{vanDam:VDVZ, Zakharov:VDVZ}, is
inconsistent, as the graviton mass goes to zero. Specifically, the
standard linearised analysis near a heavy source is only valid at distances
\be
r \geqslant \left(\frac{r_M}{m^4}\right)^{\frac{1}{5}}
\ee
where $r_M$ is the Schwarzschild radius of the source, and $m $ is the
graviton mass. To see if a
mass discontinuity really is present, we need to calculate the
Schwarzschild solution on a brane, and compare it to the standard
four-dimensional massless result. This is a highly non-trivial
exercise that is beyond the scope of this paper. 

For Pauli-Fierz theory~\cite{Fierz:pauli-fierz}, the breakdown of the
linearised analysis has been linked to strong coupling
phenomena at the following scale~\cite{Arkani-Hamed:massive}
\be
E_\textrm{strong} \sim \left(\frac{l_{pl}}{m^4}\right)^{-\frac{1}{5}}
\ee
where $l_{pl}$ is the Planck length\footnote{For certain non-linear
extensions of Pauli-Fierz, $E_\textrm{strong} \sim \left(l_{pl}/m^2\right)^{-\frac{1}{3}}$~\cite{Schwartz:decon}. This higher scale also appears in the DGP model~\cite{Luty:strong}.}.  Large terms in the full propagator tend to signal
this problem.  Unfortunately, we might
expect our theory to suffer the same fate. We can see this by looking
at $h_{zz}$ in a fixed wall gauge (see equation (\ref{fixed})). By the
mean value theorem, there exists $z_0 \in [ 0, l]$ such that $B^\prime
(z_0)=1/l$. For small $l$, it is clear that $h_{zz}$ can be very
large.

A similar strong coupling scale also  exists in the DGP
model~\cite{Luty:strong,Rubakov:strong}. However, it has recently been argued that this scale
could be unphysical, and is just a result of the naive perturbative
expansion~\cite{Dvali:IR}. Clearly, both the mass discontiniuty
and strong coupling problems are highly contentious issues at the
moment. For this reason we have focused on the possible existence of
ghosts in our models. There is no contention there: ghosts are
undesirable, but can be avoided.

There is still much to do. It would be very interesting to study the phenomenology of these models
in more detail, particularly in the context of cosmic
acceleration. Coincidentally, we have been forced to choose a brane
separation $l \sim 10^3 ~(\textrm{eV})^{-1}$, which agrees with the
brane separation chosen in~\cite{Cognola:multi}. In~\cite{Cognola:multi}, the mass spectrum is
motivated by a discretized Randall-Sundrum model, with $l$ chosen so
that there is a small effective cosmological constant. Would the
presence of an ultralight mode in the spectrum affect these results?
We would probably expect the answer to be ``no'', because the dominant
contribution to the vacuum energy would still come from the heavier
modes. Nevertheless, it is worthy of further investigation. 

Our models could certainly be generalised in a number of ways. We
could  consider more
branes, abandon $\mathbb{Z}_2$
symmetry, or even introduce higher derivative terms in the bulk and on the
brane. Branes embedded in solutions to Gauss-Bonnet gravity have been
the subject of much research recently (see, for
example~\cite{Davis:junction, Padilla:gbholog}),
motivated by the link to string theory. Indeed, it would also be nice if our
models, or at least some generalisation, could be derived from a
more fundamental theory~\cite{Corley:EH,Antoniadis:CY}. However, the high
degree of fine tuning could prove an obstacle in this respect.

Let us end by summarising our main result: we have discovered
braneworld models that exhibit bigravity, without introducing
ghosts. Recall that bigravity naturally gives rise to new
gravitational physics in the infra-red. The new physics occurs when
the massive graviton ``switches off'', at distances beyond its Compton
wavelength. Our models are an improvement on the ghost-free model
given in~\cite{Kogan:adsbranes}, because they lead to potentially
observable modifications of gravity. In~\cite{Kogan:adsbranes}, all
modifications are hidden behind the AdS horizon.

\vskip .5in
\centerline{\bf Acknowledgements}
\medskip
I would like to thank John March-Russell, Valery Rubakov, Graham Ross,
Syksy R\"as\"anen, and Ben Gripaios
for  helpful dicussions. In particular, I would like to thank John and
Graham for proof reading this article. 
Thanks must also go to Ben,
Ro, Ash and Perks for being a constant source of inspiration, Leppos
and Beyonc\'e. And to Bruno Cheyrou for being the new Zidane. AP was funded by PPARC.
\bibliographystyle{utphys}

\bibliography{multi}
\appendix \label{app}
\section{Mass spectrum for AdS branes} \label{app:spectrum}
For ``symmetric'' AdS branes, we have  
\be
a(z)=\frac{\la}{k}\cosh\left[k\left(|z|-\frac{l}{2}\right)\right],
\qquad  \la=k/\cosh\left[kl/2\right]
\ee
We want to find the massive eigenstates, $u_{m}(z)$, that satisfy equation
(\ref{eqn:specbulk}).
\be
\left[\left(\frac{m^2+2\la^2}{a^2}\right)+\frac{\del^2}{\del
z^2}-4k^2\right]u_m(z)=0
\ee
Let $u_m(z)=v_m(y)/a^2(z)$, where
$y=\tanh^2\left(k\left(|z|-\frac{l}{2}\right)\right)$. We find that
$v_m$ satisfies the following hypergeometric equation 
\be
y(1-y)v_m^{\prime\prime}(y)+\frac{1}{2}(1-7y)v_m^{\prime}(y)+\frac{m^2-4\la^2}{4\la^2}v_m(y)=0
\ee
The solution is given in terms of hypergeometric functions~\cite{Kogan:adsbranes}
\be
v_m(y)=A_mF\left(a_m, b_m, \frac{1}{2} ; y\right)+B_m\sqrt{y}F\left(a_m+ \frac{1}{2},
b_m+ \frac{1}{2},  \frac{3}{2} ; y\right)
\ee
where $A_m$, $B_m$ are arbitrary constants and
\be
a_m=\frac{5}{2}+\sqrt{\frac{9}{4}+\frac{m^2}{\la^2}}, \qquad
b_m=\frac{5}{2}-\sqrt{\frac{9}{4}+\frac{m^2}{\la^2}}
\ee
The eigenstates are therefore given by
\begin{multline} \label{eqn:adsum}
u_m(z)=\sech^2\left[k\left(|z|-\frac{l}{2}\right)\right]
\left\{A_mF\left(a_m, b_m, \frac{1}{2} ;
\tanh^2\left[k\left(|z|-\frac{l}{2}\right)\right]\right) \right. \\
\left.+B_m \tanh\left[k\left(|z|-\frac{l}{2}\right)\right]F\left(a_m +\frac{1}{2}, b_m+ \frac{1}{2}, \frac{3}{2} ; \tanh^2\left[k\left(|z|-\frac{l}{2}\right)\right]\right)\right\}
\end{multline}
where we have absorbed some constants into $A_m$ and $B_m$. After some
tedious algebra, we see that the boundary conditions
(\ref{eqn:specbc}) give the following
\ba
A_m \al_m+B_m \beta_m &=& 0 \\
A_m \tilde \al_m+B_m \tilde \beta_m &=& 0
\ea
where
\ba
\al_m &=& 4kta_mb_m\left(t^2-1 \right)    F\left(a_m+1, b_m+1,
\frac{3}{2} ;t^2 \right) \nonumber \\
&&\qquad +\left(4kt+\bar r m^2\right)F\left(a_m, b_m, \frac{1}{2}
;t^2\right)   \\ 
\beta_m &=&\left(\frac{\Delta r}{2}\right)m^2tF\left(a_m+\frac{1}{2},
b_m+\frac{1}{2}, \frac{3}{2} ; t^2\right) \\
\tilde \al_m &=& \left(\frac{\Delta r}{2}\right)m^2F\left(a_m, b_m,
\frac{1}{2} ; t^2\right) \\
\tilde \beta_m &=&  \frac{4}{3}kt^2\left(t^2-1\right)\left(a_m+\frac{1}{2} \right)\left(b_m+\frac{1}{2} \right)F\left(a_m+\frac{3}{2},
b_m+\frac{3}{2}, \frac{5}{2} ;
t^2\right)  \nonumber \\
&&\qquad + \left(k\left( 5t^2-1 \right) +\bar r m^2 t\right)F\left(a_m+\frac{1}{2},
b_m+\frac{1}{2}, \frac{3}{2} ;
t^2\right) 
\ea
Here $t=\tanh\left(kl/2 \right)$ and $\bar
r=(r_L+r_R)/2$. By demanding that there exists a non-trivial solution for $A_m$ and
$B_m$, we find a quantisation condition for $m$ 
\be \label{eqn:adsqu}
\al_m \tilde \beta_m=\tilde \al_m \beta_m
\ee

\end{document}